\documentclass[italian,english]{article}
\usepackage[T1]{fontenc}
\usepackage[latin1]{inputenc}
\usepackage{color}

\makeatletter

 \newcommand{\lyxaddress}[1]{
   \par {\raggedright #1 
   \vspace{1.4em}
   \noindent\par}
 }

\usepackage{babel}
\makeatother
\begin{document}

\title{\textbf{Gravitational waves from the $R^{-1}$ high order theory
of gravity }}

\author{\textbf{\ensuremath{}Christian Corda and \ensuremath{}Mariafelicia
De Laurentis}}

\maketitle

\lyxaddress{\begin{center}\ensuremath{}INFN - Sezione di Pisa and Universit�di Pisa, Largo Pontecorvo 3, I - 56127 PISA, Italy\end{center}}

\lyxaddress{\begin{center}\ensuremath{}Politecnico di Torino and INFN Sezione di Torino, 
Corso Duca degli Abruzzi 24, I - 10129 Torino, Italy\end{center}}

\lyxaddress{\begin{center}\textit{E-mail addresses:} \textcolor{blue}{\ensuremath{}christian.corda@ego-gw.it,
\ensuremath{}mariafelicia.delaurentist@polito.it}\end{center}}

\begin{abstract}
This paper is a review of a previous research on gravitational waves
from the $R^{-1}$ high order theory of gravity. It is shown that
a massive scalar mode of gravitational waves from the $R^{-1}$ theory
generates a longitudinal force in addition of the transverse one which
is proper of the massless gravitational waves and the response of
an arm of an interferometer to this longitudinal effect in the frame
of a local observer is computed. Important conseguences from a theoretical
point of view could arise from this approach, because it opens to
the possibility of using the signals seen from interferometers to
understand which is the correct theory of gravitation.
\end{abstract}

\lyxaddress{PACS numbers: 04.80.Nn, 04.30.Nk, 04.50.+h}

\section{Introduction}

The data analysis of interferometric gravitational waves (GWs) detectors
has recently started (for the current status of GWs interferometers
see \cite{key-1,key-2,key-3,key-4,key-5,key-6,key-7,key-8}) and the
scientific community hopes in a first direct detection of GWs in next
years. 

Detectors for GWs will be important for a better knowledge of the
Universe and also to confirm or ruling out the physical consistency
of General Relativity or of any other theory of gravitation \cite{key-9,key-10,key-11,key-12,key-13,key-14}.
This is because, in the context of Extended Theories of Gravity, some
differences between General Relativity and the others theories can
be pointed out starting by the linearized theory of gravity \cite{key-15,key-16,key-17,key-18,key-19,key-20}. 

In this paper, which is a review of a previous research on gravitational
waves from the $R^{-1}$ high order theory of gravity \cite{key-15},
it is shown that massive scalar modes of gravitational waves from
the $R^{-1}$ theory generate a longitudinal force in addition of
the transverse one which is proper of the massless gravitational waves
and the response of an arm of an interferometer to this longitudinal
effect in the frame of a local observer is computed. Important conseguences
from a theoretical point of view could arise from this approach, because
it opens to the possibility of using the signals seen from interferometers
to understand which is the correct theory of gravitation.

\section{A scalar massive mode of gravitational radiation in the $R^{-1}$
high order theory of gravity}

In \cite{key-15} it has been shown that a massive scalar mode of
gravitational radiation arises from the high order action \begin{equation}
S=\int d^{4}x\sqrt{-g}R^{-1}+\mathcal{L}_{m},\label{eq: high order 1}\end{equation}

where $R$ is the Ricci scalar curvature. Equation (\ref{eq: high order 1})
is a particular choice with respect the well known canonical one of
General Relativity (the Einstein - Hilbert action \cite{key-21,key-22})
which is 

\begin{equation}
S=\int d^{4}x\sqrt{-g}R+\mathcal{L}_{m}.\label{eq: EH}\end{equation}

From the linearizated field equations arising by the action (\ref{eq: high order 1}),
in \cite{key-15} it has been obtained a plane wave propagating in
the $z$ direction:

\begin{equation}
h_{\mu\nu}(t,z)=A^{+}(t-z)e_{\mu\nu}^{(+)}+A^{\times}(t-z)e_{\mu\nu}^{(\times)}+\Phi(t-v_{G}z)\eta_{\mu\nu}.\label{eq: perturbazione totale}\end{equation}

The term $A^{+}(t-z)e_{\mu\nu}^{(+)}+A^{\times}(t-z)e_{\mu\nu}^{(\times)}$
describes the two standard (i.e. tensorial) polarizations of gravitational
waves which arise from General Relativity, while the term $\Phi(t-v_{G}z)\eta_{\mu\nu}$
is the scalar massive field arising from the high order theory.

\section{A longitudinal force}

For a purely scalar gravitational wave eq. (\ref{eq: perturbazione totale})
can be rewritten as \cite{key-15}

\begin{equation}
h_{\mu\nu}(t-v_{G}z)=\Phi(t-v_{G}z)\eta_{\mu\nu}\label{eq: perturbazione scalare}\end{equation}
and the corrispondent line element is the conformally flat one

\begin{equation}
ds^{2}=[1+\Phi(t-v_{G}z)](-dt^{2}+dz^{2}+dx^{2}+dy^{2}).\label{eq: metrica puramente scalare}\end{equation}
But, in a laboratory environment on Earth, the coordinate system in
which the space-time is locally flat \cite{key-12,key-15,key-21,key-22}
is typically used and the distance between any two points is given
simply by the difference in their coordinates in the sense of Newtonian
physics. This frame is the proper reference frame of a local observer,
located for example in the position of the beam splitter of an interferometer.
In this frame gravitational waves manifest themself by exerting tidal
forces on the masses (the mirror and the beam-splitter in the case
of an interferometer). The effect of the gravitational wave on test
masses is described in this frame by the equation

\begin{equation}
\ddot{x^{i}}=-\widetilde{R}_{0k0}^{i}x^{k},\label{eq: deviazione geodetiche}\end{equation}
which is the equation for geodesic deviation.

But, because the linearized Riemann tensor $\widetilde{R}_{\mu\nu\rho\sigma}$
is invariant under gauge transformations \cite{key-12,key-15,key-21,key-22},
it can be directly computed from eq. (\ref{eq: perturbazione scalare}). 

From \cite{key-21} it is:

\begin{equation}
\widetilde{R}_{\mu\nu\rho\sigma}=\frac{1}{2}\{\partial_{\mu}\partial_{\beta}h_{\alpha\nu}+\partial_{\nu}\partial_{\alpha}h_{\mu\beta}-\partial_{\alpha}\partial_{\beta}h_{\mu\nu}-\partial_{\mu}\partial_{\nu}h_{\alpha\beta}\},\label{eq: riemann lineare}\end{equation}

that, in the case eq. (\ref{eq: perturbazione scalare}), begins \cite{key-15}

\begin{equation}
\widetilde{R}_{0\gamma0}^{\alpha}=\frac{1}{2}\{\partial^{\alpha}\partial_{0}\Phi\eta_{0\gamma}+\partial_{0}\partial_{\gamma}\Phi\delta_{0}^{\alpha}-\partial^{\alpha}\partial_{\gamma}\Phi\eta_{00}-\partial_{0}\partial_{0}\Phi\delta_{\gamma}^{\alpha}\}.\label{eq: riemann lin scalare}\end{equation}

The computation has been performed in details in \cite{key-15}, the
results are

\begin{equation}
\begin{array}{c}
\widetilde{R}_{010}^{1}=-\frac{1}{2}\ddot{\Phi}\\
\\\widetilde{R}_{010}^{2}=-\frac{1}{2}\ddot{\Phi}\\
\\\widetilde{R}_{030}^{3}=\frac{1}{2}m^{2}\Phi,\end{array}\label{eq: componenti riemann}\end{equation}

which show that the field is not transversal. 

Infact, using eq. (\ref{eq: deviazione geodetiche}) it results

\begin{equation}
\ddot{x}=\frac{1}{2}\ddot{\Phi}x,\label{eq: accelerazione mareale lungo x}\end{equation}

\begin{equation}
\ddot{y}=\frac{1}{2}\ddot{\Phi}y\label{eq: accelerazione mareale lungo y}\end{equation}

and 

\begin{equation}
\ddot{z}=-\frac{1}{2}m^{2}\Phi(t-v_{G}z)z.\label{eq: accelerazione mareale lungo z}\end{equation}

Then the effect of the mass is the generation of a \textit{longitudinal}
force (in addition to the transverse one). Note that in the limit
$m\rightarrow0$ the longitudinal force vanishes.

\section{The interferometer's response to the longitudinal component}

We have to recall that the scalar wave needs a frequency which falls
in the frequency-range for earth based gravitational antennas \cite{key-12,key-15},
that is the interval $10Hz\leq f\leq10KHz$ (see refs \cite{key-1,key-2,key-3,key-4,key-5,key-6,key-7,key-8}).
For a massive scalar gravitational wave, this implies \cite{key-12,key-15}

\begin{equation}
0eV\leq m\leq10^{-11}eV.\label{eq: range di massa}\end{equation}

Equations (\ref{eq: accelerazione mareale lungo x}), (\ref{eq: accelerazione mareale lungo y})
and (\ref{eq: accelerazione mareale lungo z}) give the tidal acceleration
of the test mass caused by the scalar gravitational wave respectly
in the $x$ direction, in the $y$ direction and in the $z$ direction
\cite{key-15}.

Equivalently we can say that there is a gravitational potential \cite{key-12,key-15}:

\begin{equation}
V(\overrightarrow{r},t)=-\frac{1}{4}\ddot{\Phi}(t-\frac{z}{v_{P}})[x^{2}+y^{2}]+\frac{1}{2}m^{2}\int_{0}^{z}\Phi(t-v_{G}z)ada,\label{eq:potenziale in gauge Lorentziana generalizzato}\end{equation}

which generates the tidal forces, and that the motion of the test
mass is governed by the Newtonian equation

\begin{equation}
\ddot{\overrightarrow{r}}=-\bigtriangledown V.\label{eq: Newtoniana}\end{equation}

To obtain the longitudinal component of the scalar gravitational wave
the solution of eq. (\ref{eq: accelerazione mareale lungo z}) has
to be found. 

For this goal the perturbation method can be used. A function of time
for a fixed $z$, $\psi(t-v_{G}z)$,  can be defined \cite{key-12,key-15},
for which it is

\begin{equation}
\ddot{\psi}(t-v_{G}z)\equiv\Phi(t-v_{G}z)\label{eq: definizione di psi}\end{equation}

(note: the most general definition is $\psi(t-v_{G}z)+a(t-v_{G}z)+b$,
but, assuming only small variatons in the positions of the test masses,
it results $a=b=0$).

In this way it results

\begin{equation}
\delta z(t-v_{G}z)=-\frac{1}{2}m^{2}z_{0}\psi((t-v_{G}z).\label{eq: spostamento lungo z}\end{equation}

A feature of the frame of a local observer is the coordinate dependence
of the tidal forces due by scalar gravitational waves which can be
changed with a mere shift of the origin of the coordinate system \cite{key-12,key-15}:

\begin{equation}
x\rightarrow x+x',\textrm{ }y\rightarrow y+y'\textrm{ and }z\rightarrow z+z'.\label{eq: shift coordinate}\end{equation}

The same applies to the test mass displacements in the $z$ direction,
eq. (\ref{eq: spostamento lungo z}). This is an indication that the
coordinates of a local observer are not simple as they could seem
\cite{key-12,key-15}. 

Now, let us consider the relative motion of test masses. A good way
to analyze variations in the proper distance (time) of test masses
is by means of {}``bouncing photons'' \cite{key-12,key-15}. A photon
can be launched from the beam-splitter to be bounced back by the mirror.
It will be assumed that both the beam-splitter and the mirror are
located along the $z$ axis of our coordinate system (i.e. an arm
of the interferometer is in the $z$ direction, which is the direction
of the propagating massive scalar gravitational wave and of the longitudinal
force).

In the frame of a local observer, two different effects have to be
considered in the calculation of the variation of the round-trip time
for photons \cite{key-12,key-15}. The unperturbed coordinates for
the beam-splitter and the mirror are $z_{b}=0$ and $z_{m}=L$. So
the unperturbed propagation time between the two masses is

\begin{equation}
T=L.\label{eq: tempo imperturbato}\end{equation}

From eq. (\ref{eq: spostamento lungo z}) it results that the displacements
of the two masses under the influence of the scalar gravitational
wave are

\begin{equation}
\delta z_{b}(t)=0\label{eq: spostamento beam-splitter}\end{equation}

and

\begin{equation}
\delta z_{m}(t-v_{G}L)=-\frac{1}{2}m^{2}L\psi(t-v_{G}L).\label{eq: spostamento mirror}\end{equation}

In this way, the relative displacement, is

\begin{equation}
\delta L(t)=\delta z_{m}(t-v_{G}L)-\delta z_{b}(t)=-\frac{1}{2}m^{2}L\psi(t-v_{G}L),\label{eq: spostamento relativo}\end{equation}

Thus it results

\begin{equation}
\frac{\delta L(t)}{L}=\frac{\delta T(t)}{T}=-\frac{1}{2}m^{2}\psi(t-v_{G}L).\label{eq: strain scalare}\end{equation}

But there is the problem that, for a large separation between the
test masses (in the case of Virgo or LIGO the distance between the
beam-splitter and the mirror is three or four kilometers), the definition
(\ref{eq: spostamento relativo}) for relative displacement becomes
unphysical because the two test masses are taken at the same time
and therefore cannot be in a casual connection \cite{key-12,key-15}.
The correct definitions for our bouncing photon can be written like

\begin{equation}
\delta L_{1}(t)=\delta z_{m}(t-v_{G}L)-\delta z_{b}(t-T_{1})\label{eq: corretto spostamento B.S. e M.}\end{equation}

and

\begin{equation}
\delta L_{2}(t)=\delta z_{m}(t-v_{G}L-T_{2})-\delta z_{b}(t),\label{eq: corretto spostamento B.S. e M. 2}\end{equation}
where $T_{1}$ and $T_{2}$ are the photon propagation times for the
forward and return trip correspondingly. According to the new definitions,
the displacement of one test mass is compared with the displacement
of the other at a later time to allow for finite delay from the light
propagation. Note that the propagation times $T_{1}$ and $T_{2}$
in eqs. (\ref{eq: corretto spostamento B.S. e M.}) and (\ref{eq: corretto spostamento B.S. e M. 2})
can be replaced with the nominal value $T$ because the test mass
displacements are alredy first order in $\Phi$. Thus, for the total
change in the distance between the beam splitter and the mirror in
one round-trip of the photon, it is

\begin{equation}
\delta L_{r.t.}(t)=\delta L_{1}(t-T)+\delta L_{2}(t)=2\delta z_{m}(t-v_{G}L-T)-\delta z_{b}(t)-\delta z_{b}(t-2T),\label{eq: variazione distanza propria}\end{equation}

and in terms of the amplitude and mass of the SGW:

\begin{equation}
\delta L_{r.t.}(t)=-m^{2}L\psi(t-v_{G}L-T).\label{eq: variazione distanza propria 2}\end{equation}

The change in distance (\ref{eq: variazione distanza propria 2})
leads to changes in the round-trip time for photons propagating between
the beam-splitter and the mirror:

\begin{equation}
\frac{\delta_{1}T(t)}{T}=-m^{2}\psi(t-v_{G}L-T).\label{eq: variazione tempo proprio 1}\end{equation}

In the last calculation (variations in the photon round-trip time
which come from the motion of the test masses inducted by the scalar
gravitational wave), it was implicitly assumed that the propagation
of the photon between the beam-splitter and the mirror of our interferometer
is uniform as if it were moving in a flat space-time. But the presence
of the tidal forces indicates that the space-time is curved. As a
result another effect after the previous has to be considered, which
requires spacial separation \cite{key-12,key-15}.

For this effect we consider the interval for photons propagating along
the $z$-axis

\begin{equation}
ds^{2}=g_{00}dt^{2}+dz^{2}.\label{eq: metrica osservatore locale}\end{equation}

The condition for a null trajectory ($ds=0$) gives the coordinate
velocity of the photons

\begin{equation}
v^{2}\equiv(\frac{dz}{dt})^{2}=1+2V(t,z),\label{eq: velocita' fotone in gauge locale}\end{equation}

which to first order in $\Phi$ is approximated by

\begin{equation}
v\approx\pm[1+V(t,z)],\label{eq: velocita fotone in gauge locale 2}\end{equation}

with $+$ and $-$ for the forward and return trip respectively. Knowing
the coordinate velocity of the photon, the propagation time for its
travelling between the beam-splitter and the mirror can be defined:

\begin{equation}
T_{1}(t)=\int_{z_{b}(t-T_{1})}^{z_{m}(t)}\frac{dz}{v}\label{eq:  tempo di propagazione andata gauge locale}\end{equation}

and

\begin{equation}
T_{2}(t)=\int_{z_{m}(t-T_{2})}^{z_{b}(t)}\frac{(-dz)}{v}.\label{eq:  tempo di propagazione ritorno gauge locale}\end{equation}

The calculations of these integrals would be complicated because the
boundary $z_{m}(t)$ is changing with time. In fact it is

\begin{equation}
z_{b}(t)=\delta z_{b}(t)=0\label{eq: variazione b.s. in gauge locale}\end{equation}

but

\begin{equation}
z_{m}(t)=L+\delta z_{m}(t).\label{eq: variazione specchio nin gauge locale}\end{equation}

But, to first order in $\Phi$, this contribution can be approximated
by $\delta L_{2}(t)$ (see eq. (\ref{eq: corretto spostamento B.S. e M. 2})).
Thus, the combined effect of the varying boundary is given by $\delta_{1}T(t)$
in eq. (\ref{eq: variazione tempo proprio 1}). Then, only the times
for photon propagation between the fixed boundaries $0$ and $L$
have to be calculated. Such propagation times will be denoted with
$\Delta T_{1,2}$ to distinguish from $T_{1,2}$. In the forward trip,
the propagation time between the fixed limits is

\begin{equation}
\Delta T_{1}(t)=\int_{0}^{L}\frac{dz}{v(t',z)}\approx T-\int_{0}^{L}V(t',z)dz,\label{eq:  tempo di propagazione andata  in gauge locale}\end{equation}

where $t'$ is the retardation time which corresponds to the unperturbed
photon trajectory: 

\begin{center}$t'=t-(L-z)$\end{center}

(i.e. $t$ is the time at which the photon arrives in the position
$L$, so $L-z=t-t'$).

Similiary, the propagation time in the return trip is

\begin{equation}
\Delta T_{2}(t)=T-\int_{L}^{0}V(t',z)dz,\label{eq:  tempo di propagazione andata  in gauge locale}\end{equation}

where now the retardation time is given by

\begin{center}$t'=t-z$.\end{center}

The sum of $\Delta T_{1}(t-T)$ and $\Delta T_{2}(t)$ gives the round-trip
time for photons traveling between the fixed boundaries. Then the
deviation of this round-trip time (distance) from its unperturbed
value $2T$ is

\begin{equation}
\delta_{2}T(t)=\int_{0}^{L}[V(t-2T+z,z)+V(t-z,z)]dz.\label{eq: variazione tempo proprio lungo z 2}\end{equation}

From eqs. (\ref{eq:potenziale in gauge Lorentziana generalizzato})
and (\ref{eq: variazione tempo proprio lungo z 2}) it results:

\begin{equation}
\begin{array}{c}
\delta_{2}T(t)=\frac{1}{2}m^{2}\int_{0}^{L}[\int_{0}^{z}\Phi(t-2T+a-v_{G}a)ada+\int_{0}^{z}\Phi(t-a-v_{G}a)ada]dz=\\
\\=\frac{1}{4}m^{2}\int_{0}^{L}[\Phi(t-v_{G}z-2T+z)+\Phi(t-v_{G}z-z)]z^{2}dz+\\
\\-\frac{1}{4}m^{2}\int_{0}^{L}[\int_{0}^{z}\Phi'(t-2T+a-v_{G}a)z^{2}da+\int_{0}^{z}\Phi'(t-a-v_{G}a)z^{2}da]dz,\end{array}\label{eq: variazione tempo proprio lungo z 2.2}\end{equation}

Thus the total round-trip proper distance in presence of the scalar
gravitational wave is:

\begin{equation}
T=2T+\delta_{1}T+\delta_{2}T.\label{eq: round-trip  totale in gauge locale}\end{equation}

Now, to obtain the interferometer response function of the massive
scalar field, the analysis can be transled in the frequency domine.

Using the Fourier transform of $\psi$ defined from 

\begin{equation}
\tilde{\psi}(\omega)=\int_{-\infty}^{\infty}dt\psi(t)\exp(i\omega t),\label{eq: trasformata di fourier2}\end{equation}
 and recalling a theorem about Fourier transforms \cite{key-15},
it is simple to obtain:

\begin{equation}
\tilde{\psi}(\omega)=-\frac{\tilde{\Phi}(\omega)}{\omega^{2}},\label{eq: Teorema di Fourier}\end{equation}

where

\begin{equation}
\tilde{\Phi}(\omega)=\int_{-\infty}^{\infty}dt\Phi(t)\exp(i\omega t).\label{eq: trasformata di fourier}\end{equation}

is the Fourier transform of our scalar field. Then, in the frequency
space, it results \cite{key-15}: 

\begin{equation}
\tilde{T}(\omega)=\Upsilon_{l}(\omega)\tilde{\Phi}(\omega),\label{eq: tempaccio}\end{equation}

where 

\begin{equation}
\begin{array}{c}
\Upsilon_{l}(\omega)\equiv(1-v_{G}^{2})\exp[i\omega L(1+v_{G})]+\frac{1}{2\omega L(v_{G}^{2}-1)^{2}}\\
\\{}[\exp[2i\omega L](v_{G}+1)^{3}(-2i+\omega L(v_{G}-1)+2\exp[i\omega L(1+v_{G})]\\
\\(6iv_{G}+2iv_{G}^{3}-\omega L+\omega Lv_{G}^{4})+(v_{G}+1)^{3}(-2i+\omega L(v_{G}+1))],\end{array}\label{eq: risposta totale lungo z due}\end{equation}

is the response function of an arm of our interferometer located in
the $z$-axis, due to the longitudinal component of the massive scalar
gravitational wave propagating in the same direction of the axis. 

For $v_{G}\rightarrow1$ it is $\Upsilon_{l}(\omega)\rightarrow0$.

\section{Conclusions}

In this paper, which is a review of a previous research on gravitational
waves from the $R^{-1}$ high order theory of gravity \cite{key-15},
it has been shown that massive scalar modes of gravitational waves
from the $R^{-1}$ theory generate a longitudinal force in addition
of the transverse one which is proper of the massless gravitational
waves and the response of an arm of an interferometer to this longitudinal
effect in the frame of a local observer has been computed. Important
conseguences from a theoretical point of view could arise from this
approach, because it opens to the possibility of using the signals
seen from interferometers to understand which is the correct theory
of gravitation.

\section{Acknowledgements}

We would like to thank Salvatore Capozziello for helpful advices during
the work.

\end{document}